\documentstyle[12pt,a41]{article}

\newcommand\ep{\varepsilon}

\newcommand\GeV{\,\mbox{GeV}}

\begin{document}
\setlength{\baselineskip}{0.515cm}
\sloppy
\thispagestyle{empty}
\begin{flushleft}
DESY 98--067 \hfill
{\tt hep-ph/9806357}\\
June 1998
\end{flushleft}

\setcounter{page}{0}

\mbox{}
\vspace*{\fill}
\begin{center}
{\LARGE\bf On the Gluon Regge Trajectory in \boldmath{$O(\alpha_s^2)$}}
$^{
^{\footnotemark}}$\footnotetext{Work supported in part by EU contract 
FMRX-CT98-0194.}

\vspace{5em}
\large
J. Bl\"umlein, V. Ravindran and W. L. van Neerven\footnote{On leave of
absence of Inst. Lorentz, Leiden University, The Netherlands}
\\
\vspace{5em}
\normalsize
{\it DESY--Zeuthen}\\
{\it Platanenallee 6, D--15735 Zeuthen, Germany}\\
\vspace*{\fill}
\end{center}
\begin{abstract}
\noindent
We recalculate the gluon Regge trajectory in next-to-leading order to
clarify a discrepancy between two results in the literature on the
constant part. We confirm the result obtained by Fadin et al.~\cite{FFK}.
The effects on the anomalous dimension and on the $s^{\omega}$ behavior 
of inclusive cross sections are also discussed.
\end{abstract}

\vspace{1mm}
\noindent
\begin{center}
PACS numbers~:~12.38.Cy
\end{center}

\vspace*{\fill}
%%%%%%%%%%%%%%%%%%%%%%%%%%%%%%%%%%%%%%%%%%%%%%%%%%%%%%%%%%%%%%%%%%%%%%%
\newpage
%%%%%%%%%%%%%%%%%%%%%%%%%%%%%%%%%%%%%%%%%%%%%%%%%%%%%%%%%%%%%%%%%%%%%%%%%
\section{Introduction}
\label{sectI}
%%%%%%%%%%%%%%%%%%%%%%%%%%%%%%%%%%%%%%%%%%%%%%%%%%%%%%%%%%%%%%%%%%%%%%%%%

\vspace{1mm}
\noindent
%------------------------------------------------------------------------
The Regge trajectory $\omega(L)$ forms an essential ingredient for the
kernel of the Bethe--Salpeter equation~\cite{ANOM1,ANOM3} 
to resum the small
$x$ contributions to the gluon anomalous dimension $\gamma_{gg}$ in
next-to-leading order in $\ln(1/x)$ and to all orders in the coupling
constant. $\omega(L)$ is supposed to be a process--independent quantity,
which, beyond one--loop order becomes renormalization--scheme dependent.
For definiteness we will refer to the $\overline{\rm MS}$ scheme.
There $\omega(L)$ is given by~\cite{FFK}
%-----------------------------------------------------------------------
\begin{equation}
\label{eqom}
\omega(L) = \omega_q(L) + \omega_g(L) =
- \overline{g}^2 \left[\frac{2}{\ep} + 2 L\right]
- \overline{g}^4 \left\{ c_0 \left[\frac{1}{\ep^2} -  L^2 \right]
+ c_1 \left[\frac{1}{\ep} + 2 L \right] + c_2 \right\}
\end{equation}
%-----------------------------------------------------------------------
with $\overline{g}^2 = g^2 N_c \Gamma(1-\ep)/(4\pi)^{2+\ep}$,
$L = \log(q^2/\mu^2)$, $N_c$ the number of colors, $\ep = D/2-2$, and
$D$ the dimension of space-time.
Previous
studies~\cite{FFK,KK} agree on the  the value of the coefficients
%-----------------------------------------------------------------------
\begin{eqnarray}
c_0 &=& \frac{\beta_0}{3} = \frac{11}{3} - \frac{2}{3}\frac{N_f}{N_c}  \\
c_1 &=& \frac{67}{9} - 2 \zeta(2) - \frac{10}{9}\frac{N_f}{N_c},
\end{eqnarray}
%-----------------------------------------------------------------------
but partly different results were reported on the coefficient $c_2$,
%-----------------------------------------------------------------------
\begin{eqnarray}
\label{c21}
c_2 &=& - \frac{440}{27} + ~2 \zeta(3)
+ \frac{56}{27}\frac{N_f}{N_c},~~~~~
\cite{FFK}\\
\label{c22}
c_2 &=& - \frac{440}{27} + 14 \zeta(3)
+ \frac{56}{27}\frac{N_f}{N_c},~~~~~
\cite{KK}~~.
\end{eqnarray}
%-----------------------------------------------------------------------
Here $N_f$ denotes the number of quark flavors.

The quarkonic contributions ($\omega_q(L)$) to the coefficients
$c_i$ $\propto N_f$ in eq.~(\ref{eqom}) were derived in ref.~\cite{Q1}.
The gluonic
contribution, $\omega_g(L)$, is represented by~\cite{FFK}
%-----------------------------------------------------------------------
\begin{eqnarray}
\omega_g(L) = \left(q^2\right)^{2 \ep} \left [A_0 (I_1 -  2J_1)
- 2(J_2 -J_3) - I_2 \right]
\end{eqnarray}
%-----------------------------------------------------------------------
with
%-----------------------------------------------------------------------
\begin{eqnarray}
A_0 = \psi(1) + 2 \psi(\ep) - \psi(1-\ep) -2\psi(1+2\ep)
+\frac{1}{2}\frac{(2+\ep)(3+\ep)}{\ep(1+2\ep)(3+2\ep)}
\end{eqnarray}
%-----------------------------------------------------------------------
and
the integrals $I_{1,2}$ and $J_{1,2,3}$ are
%-----------------------------------------------------------------------
\begin{eqnarray}
I_i &=&
 \overline{g}^4
\frac{1}{\pi^{2(1+\ep)}} \frac{1}{\Gamma^2(1-\ep)}
\int \frac{d^{2(1+\ep)} q_1 d^{2(1+\ep)} q_2
\left(q^2\right)^{2(1-\ep)}}
{q_1^2 q_2^2 (q_1-q)^2 (q_2-q)^2} a_i \\
J_i &=&
 \overline{g}^4
\frac{1}{\pi^{2(1+\ep)}} \frac{1}{\Gamma^2(1-\ep)}
\int \frac{d^{2(1+\ep)} q_1 d^{2(1+\ep)} q_2
\left(q^2\right)^{(1-2\ep)}}
{q_1^2 q_2^2 (q-q_1-q_2)^2} b_i~,
\end{eqnarray}
%-----------------------------------------------------------------------
where
$a_1 = b_1 = 1, a_2 = \log[(q_1-q_2)^2/q^2],
                     b_2 = \log[(q-q_2)^2/q^2]$ and
$b_3 = \log(q_2^2/q^2)$.
Standard techniques allow to calculate the integrals
$I_1, J_1, J_2$ and $J_3$ straightforwardly and
the results given in ref.~\cite{FFK} are confirmed.
The calculation of the integral $I_2$, to which the term $\propto
\zeta(3)$ in eqs.~(\ref{c21},\ref{c22}) can be attributed,
is much more difficult,
enforcing the use of certain approximation methods in the previous
calculation~\cite{FFK}.

It is the aim of the present paper to recalculate this integral by a
method different from that being used in \cite{FFK} to obtain a thorough
and independent check. Here we aim at using as general as possible
representations in all steps.

As was outlined in refs.~\cite{ANOM2,CONF} the term $\propto \zeta(3)$
in $c_2$ contributes to $\gamma_{gg}$ at the 3--loop level and is
essential for the small-$x$ behavior of this quantity. Since the the
gluonic anomalous dimension has not yet been calculated up to 3--loop
order, 
this coefficient is an important {\it prediction} and its value has
to be determined unambiguously.

%%%%%%%%%%%%%%%%%%%%%%%%%%%%%%%%%%%%%%%%%%%%%%%%%%%%%%%%%%%%%%%%%%%%%%%%%
\section{The integral $I_2$}
\label{sectM}
%%%%%%%%%%%%%%%%%%%%%%%%%%%%%%%%%%%%%%%%%%%%%%%%%%%%%%%%%%%%%%%%%%%%%%%%%
The integral $I_2$ is given by
%-----------------------------------------------------------------------
\begin{equation}
\label{eq10}
I_2 = \left. - \frac{\partial}{\partial \nu} 
\hat{I}_2(\nu) \right|_{\nu=0}
\end{equation}
%-----------------------------------------------------------------------
with\footnote{Already a long time ago integrals of the type (\ref{Imain})
were studied, representing them by doubly infinite series~\cite{TV}.
We refer, complementary to this, to integral representations in the
present treatment, since at the parameters relevant for the present
problem we found it easier to derive the singularity structure in $\ep$.}
%-----------------------------------------------------------------------
\begin{equation}
\label{Imain}
\hat{I}_2 = \overline{g}^4
\frac{1}{\pi^{2(1+\ep)}} \frac{1}{\Gamma^2(1-\ep)}
\int \frac{d^{2(1+\ep)} q_1 d^{2(1+\ep)} q_2
\left(q^2\right)^{2(1+\ep) +\nu}}
{q_1^2 q_2^2 (q_1-q)^2 (q_2-q)^2 \left[(q_1 - q_2)^2\right]^{\nu}}~.
\end{equation}
%-----------------------------------------------------------------------
A general class of integrals, to which (\ref{Imain}) belongs in
principle, was studied in ref.~\cite{BGK} recently. As will be shown 
below
the integral $I_2$ cannot be directly obtained from the solution given 
in
\cite{BGK}, but appropriate limits are required
in a series of parameters to
obtain the final result. These may be performed without leaving general
representations.

In the notation of \cite{BGK} $\hat{I}_2$ reads
%-----------------------------------------------------------------------
\begin{equation}
\hat{I}_2 = \frac{\overline{g}^4}{\Gamma^2(1-\ep)}
I_4(1,1,\nu,\delta)
\end{equation}
%-----------------------------------------------------------------------
where $I_4$ can be found in \cite{BGK} and
$\delta = \nu + 1 -\ep$. Therefore the integral above
is represented by
%-----------------------------------------------------------------------
\begin{equation}
\label{I2S}
\hat{I}_2 = \overline{g}^4 \frac{2}{2\ep-1}
\frac{\nu (\nu+1-\ep)}{\Gamma^2(1-\ep)}
G_2(1,2+\nu-\ep) G_2(1,\nu+1)
\lim_{b \rightarrow 0} S(\ep-1,b,2 \ep- \nu -1, \nu - \ep)~.
\end{equation}
%-----------------------------------------------------------------------
Here the function $G_2$ is given by
%-----------------------------------------------------------------------
\begin{equation}
\label{eqG2}
G_2(\alpha_1,\alpha_2) = G_1(\alpha_1) G_1(\alpha_2) G_1(2-2\ep-\alpha_1
- \alpha_2)
\end{equation}
%-----------------------------------------------------------------------
and
%-----------------------------------------------------------------------
\begin{equation}
G_1(\alpha) = \frac{\Gamma(1+\ep-\alpha)}{\Gamma(\alpha)}.
\end{equation}
%-----------------------------------------------------------------------
The function $S(a,b,c,d)$ in eq.~(\ref{I2S}) consists of two
terms
%-----------------------------------------------------------------------
\begin{equation}
S(a,b,c,d) = S_1(a,b,c,d) + S_2(a,b,c,d)
\end{equation}
%-----------------------------------------------------------------------
with
%-----------------------------------------------------------------------
\begin{eqnarray}
S_1(a,b,c,d) &=& \frac{\pi \cot(\pi c)}{H(a,b,c,d)} - \frac{1}{c}
\label{eqS1}
\\
S_2(a,b,c,d) &=& - \frac{b+c}{bc} F(a+c,-b,-c,b+d),
\label{eqS2}
\end{eqnarray}
%-----------------------------------------------------------------------
where
%-----------------------------------------------------------------------
\begin{equation}
\label{I2H}
H(a,b,c,d) = \frac{\Gamma(1+a) \Gamma(1+b) \Gamma(1+c) \Gamma(1+d)
\Gamma(1+a+b+c+d)}{\Gamma(1+a+c) \Gamma(1+a+d) \Gamma(1+b+c)
\Gamma(1+b+d)}~.
\end{equation}
%-----------------------------------------------------------------------
Note that the limit $\lim_{b \rightarrow 0} S_1(a,b,c,d)$ is regular.
Due to the factor $1/\Gamma^2(1-\ep)$ in eq.~(\ref{Imain}) all
$\Gamma$-functions combine into Beta-functions and rational functions
in $\ep$ and $\nu$.
This is also true for the $\Gamma$-functions in (\ref{I2H}),
guaranteeing the absence of the Mascheroni number $\gamma_E$ in the
subsequent evaluation.

The function $F(\hat{a},\hat{b},\hat{c},\hat{d})$ has the serial
representation
%-----------------------------------------------------------------------
\begin{equation}
F(\hat{a},\hat{b},\hat{c},\hat{d})
= \sum_{n=1}^{\infty} \frac{(-\hat{a})_n (-\hat{b})_n}{(1+\hat{c})_n
(1+\hat{d})_n}
\end{equation}
%-----------------------------------------------------------------------
with $(r)_n = \prod_{k=1}^n (r+k-1)$ the Pochhammer--Barnes symbol.
Since $F(\hat{a},b,\hat{c},\hat{d})$ emerges only in the limit
%-----------------------------------------------------------------------
\begin{center}
\[ - \lim_{b \rightarrow 0} \frac{1}{b} F(\hat{a},-b,\hat{c},\hat{d})
\]
\end{center}
%-----------------------------------------------------------------------
we may represent $S_2$ directly by
%-----------------------------------------------------------------------
\begin{equation}
\label{I3F2}
\lim_{b \rightarrow 0} S_2(\hat{a},b,\hat{c},\hat{d})
= \frac{\nu + 2 - 3\ep}{(2 + \nu - 2 \ep)(1 + \nu - \ep)}
{_3F_2} \left[\begin{array}{lll} 3(1-\ep)+\nu,&1,&1\\
                                 3+\nu-2\ep,&2+\nu-\ep& \end{array}
;~1 \right]~.
\end{equation}
%-----------------------------------------------------------------------
The generalized hypergeometric function $_3F_2$ of unity  argument
in (\ref{I3F2}) obeys the integral representation, cf.~\cite{TAB},
%-----------------------------------------------------------------------
\begin{eqnarray}
\lefteqn{{_3F_2} \left[\begin{array}{lll} 3(1-\ep)+\nu,&1,&1\\
                                 3+\nu-2\ep,&2+\nu-\ep& \end{array}
;~1 \right]}  \nonumber\\
& &~~~~~~~~~~~~~~= \frac{1}{B(1,1+\nu-\ep)} \int_0^1 dx (1-x)^{\nu-\ep}
{_2F_1}[3(1-\ep)+\nu, 1, 3+\nu-2\ep; x]
\end{eqnarray}
%-----------------------------------------------------------------------
and the hypergeometric function $_2F_1$ is given by
%-----------------------------------------------------------------------
\begin{equation}
{_2F_1}[3(1-\ep)+\nu, 1, 3+\nu-2\ep; x]
= \frac{1}{B(1,2+\nu-2\ep)}\int_0^1 dt (1-t)^{1+\nu-2\ep}
(1-tx)^{-3(1-\ep)-\nu}~.
\end{equation}
%-----------------------------------------------------------------------
Here $B(\alpha_1,\alpha_2)$ denotes  Euler's
Beta-function.

Therefore (\ref{I3F2}) obeys the representation
%-----------------------------------------------------------------------
\begin{eqnarray}
\label{INT2}
\lim_{b \rightarrow 0} \nu S_2
&=&  (\nu+2-3\ep) \nu  \int_0^1 dx \int_0^1 dt (1-x)^{\nu-\ep}
     (1-t)^{1+\nu-2\ep} (1-tx)^{-3(1-\ep)-\nu}
\nonumber\\
&=& - (2+\nu-3\ep) \int_0^1 dx \int_0^1 dz \left[\frac{d}{dx} (1-x)^{\nu}
\right] z^{1+\nu-2\ep} (1 -zx)^{-\ep}~.
\end{eqnarray}
%-----------------------------------------------------------------------
The factor $\nu$ in the r.h.s. of (\ref{INT2}) has to be absorbed
into the integral to obtain finite expressions in the subsequent
calculation. All the above relations are exact transformations of
$\hat{I}_2(\nu)$.

We now determine its expansion coefficient $\propto  \nu$. For this
purpose $\hat{I}_2$ is represented by
%-----------------------------------------------------------------------
\begin{eqnarray}
\label{eq25}
\hat{I}_2(\nu) &=& \overline{g}^4
C(\nu) \left[\nu S_2(\nu)\right]
= \overline{g}^4
\left[C_0 + \nu C_1\right] \left[(S_{10} + S_{20})
+ \nu (S_{11} + S_{21}) \right] + O(\nu^2)~.
\end{eqnarray}
%-----------------------------------------------------------------------
The coefficients $C_{0}, C_1, S_{10}$ and $S_{11}$ are
obtained from a serial expansion of
eqs.~(\ref{I2S}, \ref{eqG2}) and (\ref{eqS1}). Also the coefficient
$S_{20}$ can be given in a closed form by partial integration of
eq.~(\ref{INT2}). To determine $S_{21}$ we apply a Taylor expansion
of the latter integral with respect to $\nu$ and $\ep$. The
individual, partly transcendental integrals $T_i$ contributing to this
expression are listed in the appendix.

The different expansion coefficients read~:
%-----------------------------------------------------------------------
\begin{eqnarray}
C_0 &=& - \frac{6}{\ep^2} + \frac{6}{\ep} + 12 \left[1 + \zeta(2)\right]
+ 12 \left[2 - \zeta(2)- 5 \zeta(3) \right] \ep + O(\ep^2)
\\
C_1 &=&  - \frac{4}{\ep^3} - \frac{2}{\ep^2} - \frac{4}{\ep}
\left[1 - 2 \zeta(2) \right]
- 4 \left[2 - \zeta(2) - 8 \zeta(3) \right] + O(\ep)
\\
S_{10} &=&  \frac{1}{3} + \frac{1}{3} \ep +\left[1 - 2 \zeta(2) \right]
\ep^2 + \left[3 - 2 \zeta(2) - 2 \zeta(3) \right] \ep^3  + O(\ep^4)
\\
S_{11} &=&  \frac{5}{18} \frac{1}{\ep} + \frac{17}{18} + \frac{3}{2} \ep
+ \left[ \frac{3}{2} + 2 \zeta(2) + \frac{4}{3} \zeta(3) \right] \ep^2
+ O(\ep^3)
\\
S_{20} &=&  -(2-3\ep) B(2-2\ep,1-\ep) \nonumber \\
       &=&  -1 -  \ep - \left[3 - 2 \zeta(2) \right]
            \ep^2 - \left[9 - 2 \zeta(2) - 6 \zeta(3)
            \right] \ep^3  + O(\ep^4)
\\
S_{21} &=&  \frac{\ep}{4} + \frac{\ep^2}{2}
- \left[T_1 + 2 T_2  + 2 T_3\right] \ep
        +  \left[3 T_2 + 5 T_3 + T_4 + 4 T_5 + 4 T_6 + 2 T_7 + 2 T_8
           \right] \ep^2 \nonumber\\ & &
+ O(\ep^3)
\nonumber\\ &=& \frac{5}{2} \ep + \left[12 - 2 \zeta(2) - 6 \zeta(3)
\right] \ep^2
+ O(\ep^3)
\label{T22}
\end{eqnarray}
%-----------------------------------------------------------------------
leading to
%-----------------------------------------------------------------------
\begin{eqnarray}
\label{FIN}
I_2 = - \overline{g}^4
\left[ \frac{1}{\ep^3} -
\frac{2}{\ep} \zeta(2)
-26 \zeta(3) \right ]
\end{eqnarray}
%-----------------------------------------------------------------------
using eqs.~(\ref{eq10},\ref{eq25}), which confirms the result derived by
Fadin et al. in ref.~\cite{FFK} before.
Here the coefficients of $O(1/\ep^3)$ and $O(1/\ep^2)$
are independent of the value of $S_{21}$. For the constant term
$O(\ep^0)$ all rationals and the
$\zeta(2)$-terms are canceling and only $\zeta(3)$-terms contribute.
Finally $c_2$ is given by eq.~(\ref{c21}).
%%%%%%%%%%%%%%%%%%%%%%%%%%%%%%%%%%%%%%%%%%%%%%%%%%%%%%%%%%%%%%%%%%%%%%%%%
\section{Consequences for the
small-\boldmath{$x$} resummed anomalous
dimension \boldmath{$\gamma_{gg}^{NLO}$}}
\label{sectA}
%%%%%%%%%%%%%%%%%%%%%%%%%%%%%%%%%%%%%%%%%%%%%%%%%%%%%%%%%%%%%%%%%%%%%%%%%

\vspace{1mm}
\noindent
%------------------------------------------------------------------------
It is convenient to study $\Gamma_k^n$ using DIS-schemes, cf.~\cite{CIA,
ANOM2}. 
The general expression for the resummed
gluonic
small-$x$ anomalous dimension was derived up to next-to-leading order in
ref.~\cite{CIA} and may be written as
%------------------------------------------------------------------------
\begin{eqnarray}
\label{grep}
\gamma^{\rm DIS}_{gg} &=&
\gamma_{gg}^{(0)} - \alpha_s \frac{\chi_1(\gamma)
-2 \chi_0(\gamma) \chi_0'(\gamma)}{\chi_0'(\gamma)} \nonumber\\
& & + \frac{\beta_0}{4\pi} \alpha_s^2 \frac{d \log R(\gamma)}{d \alpha_s}
+\frac{4}{9}\left[1 - R(\gamma)\right] \gamma_{qg}^{Q_0}(\gamma)
 + \frac{\beta_0}{4 \pi} \alpha_s^2 \frac{d \log\left[\gamma
\sqrt{-\chi_0'(\gamma)}\right]}{d \alpha_s}~.
\end{eqnarray}
%------------------------------------------------------------------------
Here
%------------------------------------------------------------------------
\begin{eqnarray}
\chi_0(\gamma) &=& 2 \psi(1) - \psi(\gamma) - \psi(1-\gamma)~,
\end{eqnarray}
%------------------------------------------------------------------------
cf.~\cite{BFKL},
and $\gamma_{gg}^{(0)}$ is the solution of the equation
%------------------------------------------------------------------------
\begin{eqnarray}
1 = \frac{N_c \alpha_s}{\pi} \frac{1}{N-1}
\chi_0(\gamma_{gg}^{(0)})~~~~~~{\rm with}~~~~~\gamma_{gg}^{(0)}
\rightarrow \frac{N_c \alpha_s}{\pi}\frac{1}{N-1}~~~{\rm for}~~~
N \rightarrow \infty~.
\end{eqnarray}
%------------------------------------------------------------------------
The function $R(\gamma)$ is specific for the 
$\overline{\rm MS}-{\rm DIS}$ scheme~\cite{CH}
and $\gamma_{qg}^{(Q_0)}$ denotes the quark-gluon anomalous
dimension
in the $Q_0$ scheme, see ref.~\cite{CIA}. The latter quantities 
contribute only beyond 3--loop order.
For later use we decompose $\chi_1(\gamma)$ (\cite{ANOM1,ANOM3})
into the parts
%------------------------------------------------------------------------
\begin{eqnarray}
\label{eqchi1}
\chi_1(\gamma)   =
  \chi_1^{\rm scal}(\gamma)
+ \chi_1^{\rm conf}(\gamma)
\end{eqnarray}
%------------------------------------------------------------------------
where
%------------------------------------------------------------------------
\begin{eqnarray}
  \chi_1^{\rm scal}(\gamma)  &=&
\left[ \frac{\beta_0}{6} + \frac{d}{d \gamma}
\right] \left [ \chi_0^2(\gamma) + \chi_0'(\gamma) \right]~.
\end{eqnarray}
%------------------------------------------------------------------------
$\chi_0(\gamma) \equiv \chi_0^{\rm conf}(\gamma)$ and
$\chi_1^{\rm conf}(\gamma)$ are symmetric under the interchange
$\gamma \leftrightarrow 1 - \gamma$.

Both terms are called  conformal since they do not depend on the choice
of scales, cf. also~\cite{ANOM3}, and asymptotic scale invariance implies
asymptotic conformal invariance~\cite{SP}.
Retaining only these parts of the kernel of the
Bethe--Salpeter equation in refs.~\cite{BFKL,ANOM1} one obtains the
solution
%------------------------------------------------------------------------
\begin{equation}
E_k^n(\mu^2) = E_k^n(\mu_0^2) \left(\frac{\mu^2}{\mu_0^2}\right)
^{\Gamma_k^n}
\end{equation}
%------------------------------------------------------------------------
which is {\it also} the solution of the renormalization group equation
%------------------------------------------------------------------------
\begin{equation}
\left[\mu \frac{\partial}{\partial \mu} + \beta \frac{\partial}{\partial
g} + \gamma_m \frac{\partial}{\partial m} + \gamma_{O_k}
- n \gamma_{\Phi} \right] E_k^n =0
\end{equation}
%------------------------------------------------------------------------
for $m = 0$ and $\beta =0$ if one identifies
%------------------------------------------------------------------------
\begin{equation}
\Gamma_k^n  =
\gamma_{O_k} - n \gamma_{\Phi}
\equiv \sum_{l=1}^{\infty} \alpha_s^l \gamma_l^{k,n}~.
\end{equation}
%------------------------------------------------------------------------
The coefficients $\gamma_l^{k,n}$ are then the expansion coefficients
of the conformal part of (\ref{grep}).

At the one and two-loop level the contributions to $\Gamma_k^n$
result from either $\chi_0(\gamma)$ or $\chi_1^{\rm conf}(\gamma)$
only and agree with the corresponding fixed order results. Moreover
the type of DIS-scheme is unimportant at that level.

Beginning with the 3--loop order the first term in eq.~(\ref{eqchi1}),
containing the scale dependent contributions, starts to determine  the
small $x$ part of the anomalous dimension. At that level still the
specific way in which the  DIS--scheme was introduced is unimportant,
cf.~\cite{ANOM2,CONF}.
Only beginning with the 4--loop level the small-$x$ part of the
anomalous dimension starts to depend on the specific type of 
DIS-schemes. Further running coupling effects beyond those considered 
in
\cite{ANOM1} are important, cf.~\cite{CIA,ANOM2}.
The difference between eq.~(\ref{c21}) and (\ref{c22}) effects the 
gluon
anomalous dimension at the 3--loop level.
We note that
the expansion coefficient in $O(\alpha_s^3)$ of the combination
$\gamma_+ = \gamma_{gg} + (4/9) \gamma_{qg}$ would have been
larger by a factor
of 2.123 for $N_f =4$
if one would have used eq.~(\ref{c22}) instead of 
eq.~(\ref{c21}),
having thus significant impact. This factor is widely independent of
$N_f$.

We finally would like to comment upon the intercept $\omega$ describing
the
$s^{\omega}$--behavior of inclusive cross sections.
In the saddle-point approximation~\cite{ANOM1} one obtains
%------------------------------------------------------------------------
\begin{equation}
\omega_{(1)} = \omega_{(0)} \left\{ 1 + \frac{\alpha_s^2}
{\omega_{(0)}}\left[
\chi_1^{\rm scal}(1/2) +
\chi_1^{\rm conf}(1/2)\right] \right\},~~~~{\rm with}~~~~\omega_{(0)} =
12 N_c \log(2) \frac{\alpha_s}{3 \pi}
\end{equation}
%------------------------------------------------------------------------
and (cf.~\cite{ANOM1})
%------------------------------------------------------------------------
\begin{eqnarray}
\chi_1^{\rm scal}(1/2) &=& - \omega_{(0)} \left[
2\left(\frac{11}{3} - \frac{2 N_f}{3 N_c}\right)
\log(2) + 7 \frac{\zeta(3)}{\log(2)} \right] \\
\chi_1^{\rm conf}(1/2) &=& - \omega_{(0)} \left\{
- \frac{67}{9} + 2 \zeta(2) + \frac{10}{9}
\frac{N_f}{N_c} \right. \\
&+& \left.
\frac{1}{4 \log(2)} \left [ 16 \int_0^1 \frac{dx}{x}
\arctan(\sqrt{x}) \log(1/(1-x)) - 6 \zeta(3) + \frac{\pi^3}{2}
\left(\frac{27}{16} + \frac{11}{16}\frac{N_f}{N_c^3} \right)\right]
\right\}
\nonumber
\end{eqnarray}
%------------------------------------------------------------------------
corresponding to the decomposition in eq.~(\ref{eqchi1}).
Using  the full expression for
$\chi_1(\gamma = 1/2)$ in the calculation of $\omega$
one obtains~\footnote{Note
a numerical error in eq.~(16) of ref.~\cite{ANOM1}.} 
%------------------------------------------------------------------------
\begin{equation}
\label{omstrai}
\omega = 2.64762 \alpha_s(1-6.36402 \alpha_s),
\end{equation}
%------------------------------------------------------------------------
with a maximal value of 0.1040 at $Q^2 = 8.7 \cdot 10^6 \GeV^2$.
This value is only little above the soft pomeron's intercept 0.0808
and, moreover, {\it negative} for $Q^2 < 588 \GeV^2$.
Contrary to this, the conformal part of $\chi_1(\gamma)$ implies
%------------------------------------------------------------------------
\begin{equation}
\label{omco}
\omega_{\rm conf} = 2.64762 \alpha_s(1-2.54664 \alpha_s),
\end{equation}
%------------------------------------------------------------------------
with a maximal value of 0.259915 at $Q^2 = 87 \GeV^2$. Unlike the value
of $\omega$ in (\ref{omstrai})
$\omega_{\rm conf}$
is positive for $Q^2 > 1.86 \GeV^2$. Thus the conformal part shows
a reasonable behavior, while the scale--dependent effects have still 
to be understood.
If instead of eq.~(\ref{c22}), eq.~(\ref{c21})
would have been used in eqs.~(\ref{omstrai},
\ref{omco}) the coefficient inside the brackets would be smaller by
0.6210 and lead to $\omega_{\rm max} = 0.09476$ and 0.20896 respectively.
This difference, particularly in view of the last number, is not
negligible.

In summary, we confirm the results of Fadin et al. given in 
ref.~\cite{FFK}. The difference between a result reported in 
ref.~\cite{KK} and \cite{FFK} was found to have significant impact
both on the value of $\gamma_+$ in $O(\alpha_s^3)$, widely independent of
the number of flavors, and on the conformal part of the intercept
$\omega_{\rm conf}$ in the $s^{\omega}$ behaviour in next-to-leading
order.

%%%%%%%%%%%%%%%%%%%%%%%%%%%%%%%%%%%%%%%%%%%%%%%%%%%%%%%%%%%%%%%%%%%%%%%%%
\section{Appendix}
\label{sectD}
%%%%%%%%%%%%%%%%%%%%%%%%%%%%%%%%%%%%%%%%%%%%%%%%%%%%%%%%%%%%%%%%%%%%%%%%%

\vspace{1mm}
\noindent
%------------------------------------------------------------------------
The coefficients $T_i$ in eq.~(\ref{T22}) are given by
\begin{eqnarray}
%------------------------------------------------------------------------
T_1 &=& \int_0^1 dx \int_0^1 dz \frac{z^2}{1-zx} = \frac{3}{4}
\\ [2mm]
%------------------------------------------------------------------------
T_2 &=& \int_0^1 dx \int_0^1 dz \frac{z^2 \log(1-x)}{1-zx}
= - \frac{1}{2} - \frac{1}{2} \zeta(2)
\\ [2mm]
%------------------------------------------------------------------------
T_3 &=& \int_0^1 dx \int_0^1 dz \frac{z^2 \log(z)}{1-zx}  =
-1 + \frac{1}{2} \zeta(2)
\\ [2mm]
%------------------------------------------------------------------------
T_4 &=&
\int_0^1 dx \int_0^1 dz \frac{z^2 \log(1-zx)}{1-zx}  = - \frac{7}{8}
\\ [2mm]
%------------------------------------------------------------------------
T_5 &=&
\int_0^1 dx \int_0^1 dz \frac{z^2 \log(z) \log(1-x)}{1-zx}
= \frac{5}{4} - \frac{1}{4} \zeta(2) - \frac{1}{2} \zeta(3)
\\ [2mm]
%------------------------------------------------------------------------
T_6 &=&
\int_0^1 dx \int_0^1 dz \frac{z^2 \log^2(z)}{1-zx} =
 \frac{17}{8} - \frac{1}{2} \zeta(2) - \zeta(3)
\\ [2mm]
%------------------------------------------------------------------------
T_7 &=&
\int_0^1 dx \int_0^1 dz \frac{z^2 \log(z) \log(1-zx)}{1-zx} =
 \frac{31}{16} - \frac{3}{4} \zeta(2) - \frac{1}{2} \zeta(3)
\\ [2mm]
%------------------------------------------------------------------------
T_8 &=&
\int_0^1 dx \int_0^1 dz \frac{z^2 \log(1-x) \log(1-zx)}{1-zx}  =
 \frac{3}{4} + \frac{3}{4} \zeta(2) + \frac{1}{2} \zeta(3)~.
\end{eqnarray}
%------------------------------------------------------------------------

\vspace{2mm}
\noindent
{\bf Acknowledgement.}~~For a discussion we would like to thank
O.V.~Tarasov.

%%%%%%%%%%%%%%%%%%%%%%%%%%%%%%%%%%%%%%%%%%%%%%%%%%%%%%%%%%%%%%%%%%%%%%%%%

%%%%%%%%%%%%%%%%%%%%%%%%%%%%%%%%%%%%%%%%%%%%%%%%%%%%%%%%%%%%%%%%%%%%%%%%%
\end{document}